\documentclass[12pt]{article}

\PassOptionsToPackage{hidelinks,unicode,hypertexnames=false}{hyperref}

\usepackage[margin=1in]{geometry}
\usepackage[utf8]{inputenc}
\usepackage[T1]{fontenc}
\usepackage{booktabs}
\usepackage{graphicx}
\usepackage{amsmath}
\usepackage{tabularx}
\usepackage{array}
\usepackage{longtable}
\usepackage{float}
\usepackage{nameref}
\usepackage{hyperref}
\usepackage{silence}
\newcolumntype{Y}{>{\raggedright\arraybackslash}X}

\hypersetup{
  colorlinks=true,
  linkcolor=black,
  citecolor=black,
  urlcolor=black
}

\newcommand{\seeref}[1]{\hyperref[#1]{\nameref{#1}}}

\title{Damping by design in a bilateral electromechanical transducer}
\author{
Angelo Maria Sabatini\\
The BioRobotics Institute\\
Scuola Superiore Sant'Anna\\
Pisa, Italy\\
\texttt{angelo.sabatini@santannapisa.it}
}
\date{}

\begin{document}

\maketitle

\begin{abstract}
In a bilateral electromechanical transducer, energy exchange across the electromechanical interface allows the electrical termination to modify the mechanical response through electromagnetic feedback. The paper develops this idea from a simple magnetic pendulum, describing the electromechanical coupling through the flux linkage as a function of rotation angle. Torque and back electromotive force then follow from the same coupling function, making the bilateral nature of the energy conversion explicit. A radial-field moving-coil system provides a particularly simple realization in which the coupling becomes independent of angular position and the resulting electromagnetic damping is controlled by the electrical termination. A two-port representation then identifies this damping as an impedance-reflection effect and provides a quantitative criterion for separating the electrical and mechanical time scales. A numerical example compares ideal current drive with critical electrical termination and confirms the reduced second-order description against the full third-order dynamics. The resulting treatment provides a systematic and physically transparent route from familiar undergraduate electromagnetism and oscillator dynamics to electromechanical feedback, bilateral transduction, and the effect of loading in coupled physical systems.

\end{abstract}

\section{Introduction}

When a moving-coil galvanometer is connected to a voltage source, its pointer can settle into a well-damped equilibrium. If we replace that source with an ideal current source, with the galvanometer untouched, the electromagnetic damping disappears, leaving only the intrinsic damping, if any. Nothing about the galvanometer itself has changed; only the electrical impedance of the source that drives it. The electromagnetic damping coefficient of the instrument depends on the total resistance seen by the coil, source included, and it is this damping contribution that progressively disappears as the source impedance grows. In traditional instrument design, this dependence is well known: the designer specifies a target damping and selects the external resistance that achieves it \cite{doebelin}.

The moving-coil galvanometer provides a particularly simple physical system in which to examine this dependence. Its basic design, commonly referred to as the d'Arsonval movement, was patented by Jacques d'Arsonval and Marcel Deprez in 1881. At about the same time, Edward Weston developed his own movement, which he announced in 1884 \cite{weston_galvanometer}. In the decades that followed, the moving-coil galvanometer became one of the defining needle-and-scale technologies of electrical measurement. For the present purpose, however, its interest lies less in its historical role than in the transparency with which it exposes the underlying electromechanical coupling. The same motion that is produced by electromagnetic torque also generates a back electromotive force, allowing energy to flow bidirectionally across the electromechanical interface. The mechanical response can therefore depend on the electrical circuit to which the coil is connected. The moving-coil galvanometer, referred to hereafter as the d'Arsonval galvanometer, thus provides a compact example of a physical system in which electrical and mechanical domains coexist and necessarily interact.

During the nineteenth century, Maxwell, among others, studied mathematical analogies among different physical phenomena; he also investigated in depth the physical reasons for their existence. Analogy in this context means that physically different systems can be described by equations of the same mathematical form. By the beginning of the twentieth century, these ideas had developed into systematic analogies between electrical and mechanical systems \cite{maxwell_heaviside_analogy}. The electrical--mechanical analogy was subsequently refined as the limitations of the original proposals became apparent. Firestone, in particular, introduced the mobility analogy to preserve the topology of mechanical and electrical networks \cite{firestone_mobility}, while later developments further formalized the distinction between ``through'' and ``across'' variables, which led ultimately to multidomain representations based on power exchange \cite{trent_through_across,paynter_bond_graphs}.

The main element in common to all these approaches is the recognition that the boundaries between physical domains are important. A measuring instrument, similarly to a mechanical oscillator or an electrical network, cannot always be characterized in isolation. The source or load to which it is coupled may be part of its dynamics, and neglecting that coupling may therefore be a wrong simplification. The d'Arsonval galvanometer is valuable for the present purpose precisely because it spans several domains. Its construction is simple enough to allow explicit development of the magnetic, mechanical, and electrical descriptions, yet it is rich enough that the instrument cannot be characterized apart from the circuit to which it is connected. This makes it a particularly accessible setting for examining a measuring instrument not merely as a device that yields a reading, but as a coupled physical system whose behavior depends on its electrical termination.

In standard treatments, however, these properties are usually introduced as separate topics: magnetic torque in the context of magnetism, back electromotive force in a chapter on Faraday’s law, and electromagnetic damping as a feature of instrument design. Moreover, even where derived, they are typically obtained independently of one another, rather than traced back to a common coupling principle \cite{doebelin}. The present paper shows how torque and back EMF follow from the same coupling function, obtained through a single geometric argument --- the surface swept by the coil as it rotates --- common to both a uniform-field magnetic pendulum and the radial-field d’Arsonval movement. Bilateral transduction, electromagnetic damping, and impedance reflection then emerge cleanly from this common coupling.

The treatment is intended for undergraduate students familiar with electromagnetism, damped oscillations, and elementary circuit theory. By developing the electromechanical coupling from first principles before introducing its network representation, the paper provides a systematic and physically transparent route from these familiar topics to the exchange of energy between electrical and mechanical degrees of freedom and to its consequences for the dynamics of the coupled system. The numerical example and accompanying MATLAB code also provide a directly reproducible computational exercise for classroom use, allowing students to compare the complete electromechanical dynamics with the reduced model under different electrical driving conditions.

\section{The moving-coil equation of motion}

\begin{figure}[!ht]
\centering
\includegraphics[width=0.45\columnwidth]{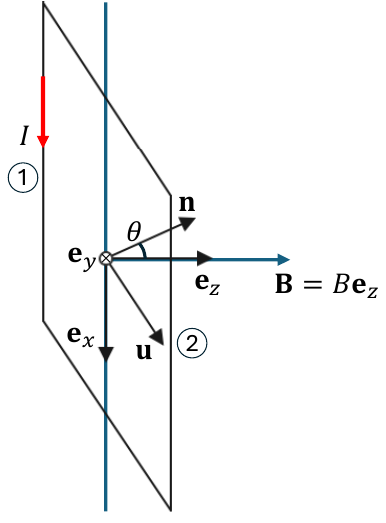}
\caption{Reference frame and loop geometry. The rotation axis coincides with $\mathbf e_x$; the uniform field is $\mathbf B=B\mathbf e_z$; the unit vector $\mathbf e_y$ points into the plane of the page, as indicated by $\otimes$. The angle $\theta$ is measured between $\mathbf e_z$ and the loop's normal $\mathbf n$; $\mathbf u$ is the in-plane vector perpendicular to the rotation axis. Sides 1 and 2 are the active sides, namely those that contribute to the change of flux and to the mechanical torque. The current flows as indicated on side 1. Schematic, not to scale.}
\label{fig-pendulum-schematic}
\end{figure}

Consider a rigid, planar coil of $N$ turns and area $A$, free to rotate (without friction) about an axis lying in its own plane, and exposed to a uniform magnetic field of magnitude $B$ perpendicular to that axis. Let $\theta$ be the angle between the unit normal to the plane of the coil, $\mathbf n$, and $\mathbf B$, and let the coil carry a current $I$, held fixed by an ideal current source (Fig.~\ref{fig-pendulum-schematic}). Gravity plays no role in determining the coil's motion: its center of mass is assumed to lie on the rotation axis, so its weight exerts no torque about that axis, whatever $\theta$ is.
The magnetic moment is $\mathbf m=NIA\,\mathbf n$, and the magnetic torque acting on it, $\boldsymbol\tau=\mathbf m\times\mathbf B$ (right-hand rule), is
\begin{equation}
\tau(\theta) = -NIAB\sin\theta.
\label{eq-torque}
\end{equation}
For a planar coil in a uniform magnetic field, it is worth noting that the shape of the coil, rectangular, circular, or otherwise, does not affect the torque once its area is fixed. A direct calculation from the Laplace force, $d\mathbf F=I\,d\boldsymbol\ell\times\mathbf B$, reaches the same result only after considerably more algebra.

The torque vanishes at $\theta=0$ and $\theta=\pi$; since $d\tau/d\theta=-NIAB\cos\theta$ is negative at $\theta=0$ and positive at $\theta=\pi$, only $\theta=0$ is a stable equilibrium. With moment of inertia $J$ about the rotation axis, the moving-coil equation of motion is
\begin{equation}
J\ddot\theta + NIAB\sin\theta = 0,
\label{eq-motion}
\end{equation}
where Newton's dot notation is used for time differentiation ($\dot x = dx/dt$; $\ddot x = d^2x/dt^2$). It is important to note that the shape invariance of the magnetic torque mentioned above does not extend to the dynamics: the moment of inertia $J$ depends on the coil geometry (as will the self-inductance $L$).

Released from rest at some $0<\theta_0<\pi$, the coil does not simply relax back to equilibrium: with no dissipation yet in the picture, it swings through $\theta=0$ and oscillates, trading what looks like kinetic energy for what looks like potential energy. We will make the bookkeeping precise next. The moving coil is therefore a magnetic pendulum and, perhaps surprisingly, its equation of motion is identical in form to that of a simple mechanical pendulum.

To see the magnetic pendulum from another angle, introduce the coil's flux linkage, $\lambda(\theta,i)=\Phi(\theta)+Li$, where $\Phi(\theta)$ is the flux linkage due to the external field $\mathbf B$ and $L$ is independent of $\theta$.

To make the geometry explicit, we first write the unit vector $\mathbf n$, normal to the plane of the coil and oriented according to the right-hand rule with respect to the current, and the unit vector $\mathbf u$, lying in the plane of the coil and perpendicular to the rotation axis, as shown in Fig.~\ref{fig-pendulum-schematic}:
\begin{equation}
\mathbf n(\theta)=\cos\theta\,\mathbf e_z-\sin\theta\,\mathbf e_y,
\qquad
\mathbf u(\theta)=\cos\theta\,\mathbf e_y+\sin\theta\,\mathbf e_z,
\label{eq-pendulum-geometry}
\end{equation}
with $\mathbf B=B\mathbf e_z$.

The two active sides, 1 and 2, each of length $\ell$, are parallel to the rotation axis $\mathbf e_x$. Their transverse positions relative to the rotation axis are
\begin{equation}
\mathbf r_1=-\frac{w}{2}\mathbf u,
\qquad
\mathbf r_2=+\frac{w}{2}\mathbf u,
\end{equation}
where $w$ is the length of the inactive sides.

Since $d\mathbf u=\mathbf n\,d\theta$, a positive rotation $d\theta$ displaces them by
\begin{equation}
d\mathbf r_1=-\frac{w}{2}\mathbf n\,d\theta,
\qquad
d\mathbf r_2=+\frac{w}{2}\mathbf n\,d\theta.
\end{equation}
The current runs along $+\mathbf e_x$ on side 1 and along $-\mathbf e_x$ on side 2. An oriented line element along either active side can therefore be written as
\begin{equation}
d\boldsymbol\ell_1=\mathbf e_x\,dx,
\qquad
d\boldsymbol\ell_2=-\mathbf e_x\,dx,
\qquad 0\leq x\leq\ell.
\end{equation}
The oriented surface elements added to the surface bounded by the coil during the rotation are therefore
\begin{equation}
d\mathbf S_j=d\mathbf r_j\times d\boldsymbol\ell_j.
\end{equation}
For the two active sides this gives
\begin{equation}
d\mathbf S_1=d\mathbf S_2 =-\frac{w}{2}\mathbf u\,dx\,d\theta.
\label{eq-swept-pendulum}
\end{equation}
Indeed, $d\mathbf r_2=-d\mathbf r_1$ and $d\boldsymbol\ell_2=-d\boldsymbol\ell_1$, so that the two sign changes cancel in their cross product. The two sides therefore contribute with the same sign, not by assumption but as a consequence of the geometry. Since $\mathbf B\cdot\mathbf u=B\sin\theta$,
\begin{equation}
d\Phi
=N\sum_{j=1}^{2}\int_0^\ell\mathbf B\cdot d\mathbf S_j
=-NAB\sin\theta\,d\theta,
\end{equation}
where $A=w\ell$. Hence
\begin{equation}
\frac{d\Phi}{d\theta}=-NAB\sin\theta,
\end{equation}
as follows independently from $\Phi(\theta)=NAB\cos\theta$.

We now introduce the magnetic coenergy $W'(i,\theta)$, an energy-like function whose derivative with respect to a mechanical coordinate, at fixed current, gives the corresponding generalized force \cite{coenergy_reference}. Here, it is convenient to write
\begin{equation}
W'(i,\theta)=\frac{1}{2}Li^2+i\Phi(\theta),
\end{equation}
so that
\begin{equation}
\tau(\theta) = \left.\dfrac{\partial W'}{\partial\theta}\right|_{i=I} = I\,\dfrac{d\Phi}{d\theta},
\label{eq-coenergy-torque}
\end{equation}
which reproduces Eq.~(\ref{eq-torque}) and provides a relation we will use again below.

At fixed current, the effective potential energy can be written in its usual magnetic-dipole form,
\begin{equation}
U(\theta)=-\mathbf m\cdot\mathbf B=-NIAB\cos\theta.
\end{equation}
Since $\Phi(\theta)=NAB\cos\theta$, the potential energy can also be written as
\begin{equation}
U(\theta)=-I\Phi(\theta).
\label{eq-potential-flux}
\end{equation}
$U(\theta)$ is defined up to an additive constant. At fixed current, the self-inductive term $\tfrac12LI^2$ is itself constant and can therefore be absorbed into this arbitrary constant (this will no longer be the case when the current is allowed to vary). Equation~(\ref{eq-motion}) then admits the first integral $\tfrac12J\dot\theta^2+U(\theta)=E$, with $E$ set by the initial conditions ($\theta(0)=\theta_0$; $\dot\theta(0)=0$). This is exactly the bookkeeping behind the oscillation of the magnetic pendulum described above.

The flux function $\Phi(\theta)$ provides a natural way to characterize the electromechanical interaction independently of the particular geometry. Define the electromechanical coupling coefficient
\begin{equation}
\kappa(\theta)=\dfrac{d\Phi}{d\theta},
\label{eq-coupling}
\end{equation}
so that the torque at fixed current takes the compact form
\begin{equation}
\tau(\theta)=\kappa(\theta)I.
\label{eq-torque-coupling}
\end{equation}
For a uniform-field geometry, $\kappa(\theta)=-NAB\sin\theta$. The coupling is thus angle dependent and vanishes at the stable equilibrium, $\theta=0$.

\begin{figure}[!ht]
\centering
\includegraphics[width=0.45\columnwidth]{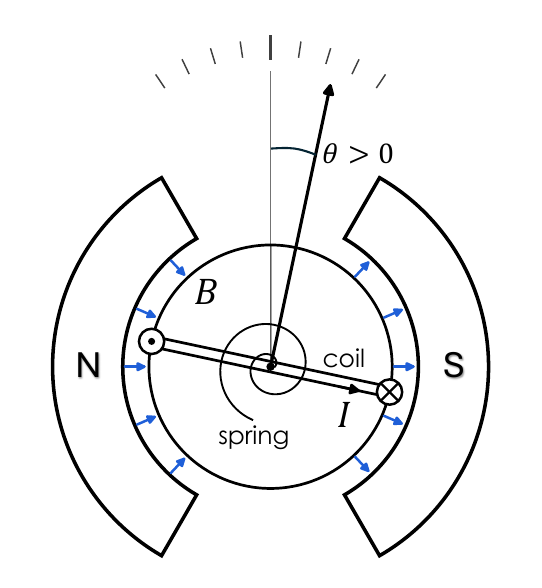}
\caption{Schematic cross section of the d'Arsonval movement, perpendicular to the rotation axis. Curved pole pieces (N, S) and a concentric soft-iron core keep the field radial in the annular air gap; the coil's two active sides at radius $a$ carry current in opposite directions ($\odot/\otimes$) and rotate together through the angle $\theta$ measured from a fixed reference. Each active side has axial length $\ell$ along the rotation axis. Schematic, not to scale.}
\label{fig-darsonval-schematic}
\end{figure}

The d'Arsonval galvanometer is built around a different choice of $\kappa(\theta)$, obtained not by reshaping the coil but by reshaping the field it turns in: curved pole pieces and a concentric soft-iron core keep the field radial throughout the coil's operating range (Fig.~\ref{fig-darsonval-schematic}).

To see the consequence, consider a stator-fixed cylindrical coordinate system $(r,\theta,z)$, with the rotation axis along $z$, and label the two active sides 1 and 2 as shown in Fig.~\ref{fig-darsonval-schematic}. The positive direction of $\theta$ is chosen to have the same sense of rotation as for the magnetic pendulum. Unlike the magnetic pendulum, however, the radial-field geometry does not define a privileged angular position, so the angular reference may be chosen arbitrarily.

The same swept-surface construction used for the magnetic pendulum now makes the effect of the radial field explicit. A positive rotation $d\theta$ makes each active side sweep a surface strip of width $a\,d\theta$. Taking into account the opposite orientations of the two active sides, the corresponding contributions to the change in flux linkage have the same sign.

The essential difference is in the scalar product with the magnetic field. For the magnetic pendulum, the uniform field is fixed in direction and its projection onto the oriented swept surface produces the factor $\sin\theta$. In the d'Arsonval geometry, the field is radial and remains normal to the surface swept by each active side throughout the rotation. Hence, with the orientation shown in Fig.~\ref{fig-darsonval-schematic},
\begin{equation}
\mathbf B\cdot d\mathbf S=B\,dS
=Ba\,d\theta\,dz
\end{equation}
for either active side. The two active sides and the $N$ turns therefore give
\begin{equation}
d\Phi
=2N\int_0^\ell \mathbf B\cdot d\mathbf S
=2Na\ell B\,d\theta.
\end{equation}
The electromechanical coupling coefficient is therefore constant over the operating range of the instrument:
\begin{equation}
\kappa(\theta)=\frac{d\Phi}{d\theta}=2Na\ell B=k.
\label{eq-darsonval-coupling}
\end{equation}

Equivalently, the flux linkage may be written
\begin{equation}
\Phi(\theta)=\Phi(\theta_{\mathrm{ref}})+k(\theta-\theta_{\mathrm{ref}}),
\end{equation}
where $\theta_{\mathrm{ref}}$ is the chosen angular reference of the instrument. In the traditional description of the d'Arsonval movement, the radial field thus produces a \emph{reduction of $\sin\theta$ to unity}. The phrase does not describe an approximation: the angle-dependent factor that appeared in the swept-surface calculation for the uniform-field magnetic pendulum is absent because the radial field remains normal to the swept surface throughout the motion.

This comes at a price. With $\kappa=k$ constant, the electromagnetic torque $kI$ is independent of $\theta$ and therefore provides no restoring action. A restoring torque must instead be supplied separately, typically by a suspension or hairspring, $\tau_{\mathrm{spring}}=-K(\theta-\theta_{\mathrm{ref}})$. With no loss of generality, we assume $\theta_{\mathrm{ref}}=0$. The equation of motion becomes
\begin{equation}
J\ddot\theta+K\theta=kI,
\label{eq-dArsonval-motion}
\end{equation}
with a static balance at $\theta_{\mathrm{steady}}=kI/K$, the reading a real d'Arsonval galvanometer produces.

\section{The introduction of damping}

Replace the ideal current source with a real one: a Norton source $I_p$ in parallel with an internal resistance $R_p$, feeding the coil through its own resistance $R_c$. Once the current is no longer imposed independently of the coil's motion, the changing flux linkage must also be included in the electrical dynamics. As the coil rotates,
\begin{equation}
\frac{d\Phi}{dt}=\frac{d\Phi}{d\theta}\dot\theta=\kappa(\theta)\dot\theta.
\label{eq-flux-rate}
\end{equation}
By Faraday--Lenz law, the resulting back electromotive force (EMF), defined here as the voltage drop opposing the motion-induced change in flux linkage, is
\begin{equation}
e_{\mathrm{b}} = \kappa(\theta)\dot\theta.
\label{eq-back-emf}
\end{equation}
Converting the Norton source to its Thévenin equivalent, $V_s=I_pR_p$, Kirchhoff's voltage law around the loop then reads
\begin{equation}
L\dot{i}+R\,i+\kappa(\theta)\dot\theta=V_s,
\label{eq-kvl}
\end{equation}
where $R=R_p+R_c$ and $\kappa(\theta)$ is the electromechanical coupling coefficient defined in Sec.~II.

For the moment, neglect the coil's electrical transient by setting $L\,\dot{i}\simeq0$ in Eq.~(\ref{eq-kvl}). The current is then \emph{slaved} to the instantaneous mechanical state $(\theta,\dot\theta)$ according to
\begin{equation}
i(\theta,\dot\theta)= I_0-\frac{\kappa(\theta)}{R}\dot\theta,\qquad I_0=\frac{V_s}{R}.
\label{eq-current}
\end{equation}
The conditions under which this quasi-static approximation is justified will be examined quantitatively below.

Substituting Eq.~(\ref{eq-current}) into the general torque relation $\tau=\kappa(\theta)\,i$ of Sec.~II separates the torque into its static-current and motion-induced contributions:
\begin{equation}
\tau = \kappa(\theta)I_0-b(\theta)\dot\theta, \qquad b(\theta)=\frac{\kappa(\theta)^2}{R}.
\label{eq-damping}
\end{equation}
The contribution of the damping term to the mechanical power is
\begin{equation}
P_{\mathrm{em}} =-b(\theta)\dot\theta^2 =-\frac{\kappa(\theta)^2}{R}\dot\theta^2\leq 0.
\label{eq-damping-power}
\end{equation}
Thus the back-action of the electrical circuit can only remove mechanical energy from the coil. Whatever coupling $\kappa(\theta)$ a design produces, its motion-induced contribution to the torque is necessarily dissipative.

For the uniform-field coil of Sec.~II, $\kappa(\theta)=-NAB\sin\theta$, and Eq.~(\ref{eq-damping}) gives the damped equation of motion
\begin{equation}
J\ddot\theta+b(\theta)\dot\theta = -NI_0AB\sin\theta, \qquad b(\theta)=\frac{\kappa(\theta)^2}{R} = \frac{(NAB\sin\theta)^2}{R}.
\label{eq-pendulum-damped}
\end{equation}
This damping, however, disappears precisely where it would be most useful: because $\kappa(0)=0$, $b(\theta)$ vanishes quadratically as $\theta\to0$. Electromagnetic damping therefore does not even appear in the linearized dynamics about the stable equilibrium; it enters only at higher order.

For the d'Arsonval coil of Sec.~II, $\kappa=k$, so the same substitution in the spring-restored equation of motion, Eq.~(\ref{eq-dArsonval-motion}), gives
\begin{equation}
J\ddot\theta+b\dot\theta+K\theta
=kI_0,
\qquad
b=\frac{k^2}{R}.
\label{eq-dArsonval-damped}
\end{equation}
This is the puzzle of Eq.~(\ref{eq-pendulum-damped}) resolved by design, not by approximation: the same reshaping of the field that made $\kappa$ constant in Sec.~II makes the electromagnetic damping coefficient constant here as well, including at $\theta=0$.

Real galvanometer movements, however, also dissipate energy through mechanisms intrinsic to their construction. Depending on the design, these may include eddy-current damping in conducting structural parts, fluid damping, and friction associated with the suspension or bearings \cite{doebelin}. At the level of the mechanical equation of motion, their combined effect may be represented by an equivalent intrinsic damping coefficient $b_0$, giving an additional torque $-b_0\dot\theta$. The d'Arsonval equation then becomes
\begin{equation}
J\ddot\theta+
\left(b_0+\frac{k^2}{R}\right)\dot\theta
+K\theta
=kI_0.
\label{eq-darsonval-intrinsic-damping}
\end{equation}
Equation~(\ref{eq-darsonval-intrinsic-damping}) is an ordinary damped harmonic oscillator. Critical damping requires
\begin{equation}
b_0+\frac{k^2}{R}=2\sqrt{JK}.
\label{eq-critical-total-damping}
\end{equation}
Since $R=R_p+R_c$, the corresponding external critical damping resistance (CDRX) is
\begin{equation}
R_{\mathrm{CDRX}} = \frac{k^2}{2\sqrt{JK}-b_0}-R_c.
\label{eq-cdrx-general}
\end{equation}
This expression assumes $b_0<2\sqrt{JK}$ and yields a physically realizable positive external resistance only when $k^2/(2\sqrt{JK}-b_0)>R_c$; otherwise critical damping cannot be obtained by adding a passive series resistance. This is the traditional external critical resistance: the resistance external to the galvanometer for which intrinsic and electromagnetic damping together produce critical damping \cite{wenner_1916_galvanometers}. For $b_0=0$, the entire damping requirement is supplied through the electrical port.

The same result also closes the loop opened in Sec.~I. Compare two sources adjusted to deliver the same steady current $I_0$ to the same coil, but having very different output resistances. In the Norton representation used above, decreasing $R_p$ moves the drive toward a low-output-resistance voltage source: the total resistance $R=R_p+R_c$ decreases and the electromagnetic damping $k^2/R$ increases. In the opposite limit, $R_p\to\infty$, the source approaches an ideal current source. Its prescribed current remains $I_0$, but the back EMF can no longer perturb that current; correspondingly, $R\to\infty$ and the electromagnetic damping disappears. Nothing about the coil, the magnet, or the spring has changed between the two limits; only the impedance of what drives it has changed.

\section{The moving coil as a bilateral transducer}

Sections II and III already relied on a single function to describe two different effects: the torque law $\tau=\kappa(\theta)\,i$ of Sec.~II and the motion-induced EMF $e_\mathrm{b}=\kappa(\theta)\dot\theta$ of Sec.~III, both built from $\kappa(\theta)=d\Phi/d\theta$. This is not a coincidence of the geometry. Since $\tau=\partial W'/\partial\theta$ and $\lambda=\partial W'/\partial i=Li+\Phi(\theta)$ for the coenergy $W'(i,\theta)=\tfrac12Li^2+i\Phi(\theta)$ of Sec.~II, Schwarz's theorem on mixed partial derivatives guarantees
\begin{equation}
\frac{\partial\tau}{\partial i}
=\frac{\partial^2W'}{\partial i\,\partial\theta}
=\frac{\partial^2W'}{\partial\theta\,\partial i}
=\frac{\partial\lambda}{\partial\theta}
=\frac{d\Phi}{d\theta}.
\end{equation}
The same coupling function therefore governs conversion in both directions: electrical current produces mechanical torque, and mechanical motion produces an electrical EMF. The opposing sign of the latter in the electrical balance follows from the chosen port orientation, consistently with Lenz's law. A device in which the same coupling coefficient governs forward and reverse conversion is a bilateral transducer, which can be modeled using a two-port network \cite{bentley_measurement_systems}.

Following standard terminology, the electrical port carries the effort/flow pair $(v,i)$, and the mechanical port the pair $(\tau,\dot\theta)$. At either port, effort times flow is power, while effort divided by flow defines an impedance, exactly as voltage divided by current does at the electrical port. For the two domains considered here, the correspondence is direct: $L$ and $J$ are inertances, $K$ is a stiffness, and electrical resistance and mechanical damping are dissipative impedances. These correspondences are those of the Maxwell analogy.

To expose the two-port structure, it is useful first to separate the moving coil from the source connected to it. An external torque $\tau$ is taken as positive when applied in the positive direction of $\theta$. For the d'Arsonval geometry, where $\kappa=k$ is constant, the coupled equations are
\begin{equation}
\begin{cases}
v = R_c i + L\dot{i} + k\dot\theta,\\[6pt]
J\ddot\theta+b_0\dot\theta+K\theta = ki+\tau.
\end{cases}
\label{eq-coupled-time}
\end{equation}

With zero initial conditions, let $\mathrm{V}$, $\mathrm{I}$, $\mathrm{T}$, $\mathrm{\Theta}$, and $\mathrm{\Omega}=s\mathrm{\Theta}$ denote the Laplace transforms of $v$, $i$, $\tau$, $\theta$, and $\dot\theta$, respectively. The dependence on $s$ will be left implicit for compactness. The coupled equations then become
\begin{equation}
\begin{bmatrix}
\mathrm{V}\\
\mathrm{T}
\end{bmatrix}
=
\begin{bmatrix}
R_c+sL & k\\
-k & Z_{mm}(s)
\end{bmatrix}
\begin{bmatrix}
\mathrm{I}\\
\mathrm{\Omega}
\end{bmatrix},
\qquad
Z_{mm}(s)=Js+b_0+\frac{K}{s}.
\label{eq-twoport}
\end{equation}
The mechanical impedance follows directly from the correspondence introduced above: $Js$ is the inertance term, $b_0$ the dissipative term, and $K/s$ the stiffness term. The off-diagonal entries, $+k$ and $-k$, have equal magnitude and opposite sign under the chosen port orientations. This antisymmetric coupling is the form that circuit theory associates with an ideal \emph{gyrator} \cite{tellegen1948gyrator}. The equality of their magnitudes is the cross-coupling relation derived above from the common coenergy function; the opposite signs ensure that the corresponding power transfer is conservative. Indeed, the two coupling terms contribute equal and opposite powers, $k\dot\theta i$ and $-ki\dot\theta$, so the ideal coupling itself neither stores nor dissipates energy.

At the mechanical port, the galvanometer's shaft carries only its own suspension and pointer in ordinary operation, with no externally applied torque. Thus $\tau=0$. In generalized network language, the mechanical port is short circuited: its effort variable vanishes while its flow variable remains free to evolve. The second row of Eq.~(\ref{eq-twoport}) then gives
\begin{equation}
\mathrm{\Omega}=\frac{k}{Z_{mm}(s)}\,\mathrm{I}.
\end{equation}
Substitution into the first row gives the electrical driving-point impedance of the movement,
\begin{equation}
Z_{ee}(s)
=
\frac{\mathrm{V}}{\mathrm{I}}
=
R_c+sL+\frac{k^2}{Z_{mm}(s)}
=
R_c+sL+\frac{k^2}{Js+b_0+K/s}.
\label{eq-Zee}
\end{equation}
The mechanical dynamics therefore appear directly at the electrical port through the reflected impedance $k^2/Z_{mm}(s)$: the coil cannot, in general, be represented by its resistance and inductance alone.

\begin{figure}[!ht]
\centering
\includegraphics[width=0.7\columnwidth]{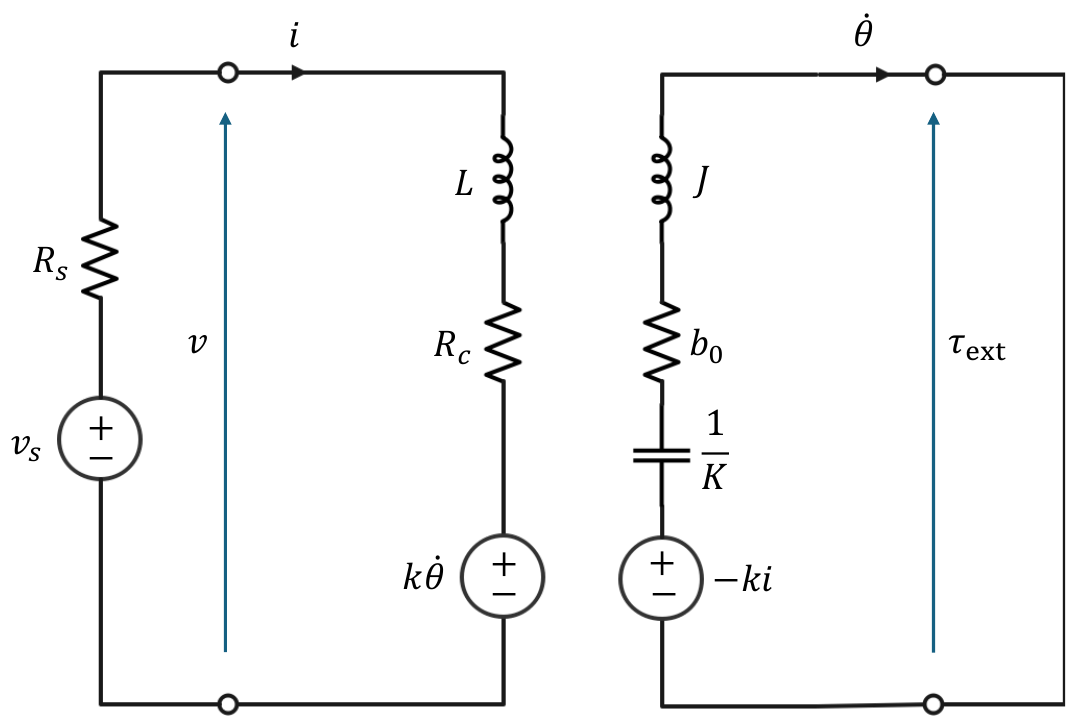}
\caption{Two-port representation of the d'Arsonval moving-coil system in the impedance analogy. The electromechanical coupling is represented by an ideal gyrator of gyration resistance $k$. The electrical port is driven by a Thévenin source of voltage $V_s$ and series resistance $R_s$, while the mechanical port is short circuited, corresponding to zero externally applied torque, $\tau=0$.}
\label{fig-twoport}
\end{figure}

We now reconnect the source of Sec.~III. Its Thévenin equivalent has voltage $V_s$ and series resistance $R_s=R_p$, where $R_p$ was the parallel resistance of the Norton representation. Figure~\ref{fig-twoport} shows the resulting electrical termination together with the short-circuited mechanical port.

The source relation is
\begin{equation}
v=V_s-R_s i.
\label{eq-source-termination}
\end{equation}
When the independent voltage source is set to zero, the electrical impedance terminating the gyrator is $R_s+R_c+sL=R+sL$, where
\begin{equation}
R=R_s+R_c.
\end{equation}
Viewed from the mechanical port, this electrical termination contributes the reflected impedance
\begin{equation}
\frac{k^2}{R+sL}.
\label{eq-reflected-mechanical}
\end{equation}
In the quasi-static limit $sL\ll R$,
\begin{equation}
\frac{k^2}{R+sL}\simeq\frac{k^2}{R}=b,
\end{equation}
recovering directly the electromagnetic damping coefficient of Sec.~III. Electromagnetic damping is therefore the electrical termination seen through the gyrator from the mechanical port, just as the mechanical dynamics appear as a reflected impedance at the electrical port in Eq.~(\ref{eq-Zee}).

This reflection also provides a quantitative criterion for the quasi-static approximation introduced in Sec.~III. The electrical relaxation time is $L/R$, whereas the natural mechanical time scale is $\sqrt{J/K}$. Neglecting the electrical transient is therefore justified when
\begin{equation}
\frac{L}{R}\ll\sqrt{\frac{J}{K}},
\label{eq-timescale-separation}
\end{equation}
or, equivalently, $\omega_nL/R\ll1$, where $\omega_n=\sqrt{K/J}$. The electrical dynamics are then fast compared with the mechanical dynamics, and the current rapidly becomes slaved to the mechanical state through the quasi-static relation used in Sec.~III. This separation of fast and slow dynamics is the standard setting of singular perturbation theory \cite{kokotovic1986}.

\section{Numerical example}

Table~\ref{tab-prototype} summarizes the parameters of the d'Arsonval galvanometer considered in the numerical example of this section.

\begin{table}[!ht]
\centering
\caption{Geometry and parameters of the example galvanometer.}
\label{tab-prototype}
\renewcommand{\arraystretch}{0.88}
\begin{tabular}{l l l}
Quantity & Symbol & Value \\
\hline
Turns & $N$ & $100$ \\
Air-gap radius & $a$ & $1\,\text{cm}$ \\
Active-side length & $\ell$ & $1.5708\,\text{cm}$ \\
Wire radius (copper) & $r$ & $50\,\mu\text{m}$ \\
Field & $B$ & $0.2\,\text{T}$ \\
\hline
Coupling constant & $k$ & $6.283\times10^{-3}\,\text{N\,m/A}$ \\
Moment of inertia & $J$ & $3.149\times10^{-8}\,\text{kg\,m}^2$ \\
Coil resistance & $R_c$ & $15.276\,\Omega$ \\
Coil inductance & $L$ & $7.271\times10^{-4}\,\text{H}$ \\
Spring constant & $K$ & $6.000\times10^{-7}\,\text{N\,m/rad}$ \\
Intrinsic damping & $b_0$ & $4.170\times10^{-8}\,\text{N\,m\,s}$ \\
\hline
External critical damping resistance & $R_{\mathrm{CDRX}}$ & $154.004\,\Omega$ \\
Total resistance at critical damping & $R$ & $169.280\,\Omega$ \\
\end{tabular}
\end{table}

The moment of inertia $J$ is calculated directly from the wire geometry, treating each side of the rectangular winding as a thin rod. The two active sides rotate at the fixed radius $a$, whereas each end side rotates about an axis through its midpoint. The coil resistance $R_c$ follows from the total wire length and cross-sectional area for the resistivity of copper. The coil inductance $L$, less readily obtained from elementary geometrical considerations, follows from two classical results for straight round wires \cite{Grover1946}.

We first consider the galvanometer driven by an ideal current source, with zero initial deflection and angular velocity. For a current step of amplitude $I_0$, the current is prescribed independently of the coil motion and therefore does not constitute a dynamical state. As discussed in Sec.~III, this corresponds to the limit $R_p\to\infty$, for which the electromagnetic damping $k^2/R$ vanishes. Equation~(\ref{eq-darsonval-intrinsic-damping}) consequently reduces to
\begin{equation}
J\ddot\theta+b_0\dot\theta+K\theta=kI_0.
\label{eq-current-driven-step}
\end{equation}
The current amplitude is chosen to give a steady deflection $\theta_{\mathrm{ss}}=30^\circ$, so that
\begin{equation}
I_0=\frac{K\theta_{\mathrm{ss}}}{k}.
\label{eq-current-step-amplitude}
\end{equation}
For the parameters of Table~\ref{tab-prototype}, $I_0=50.0\,\mu\mathrm{A}$. The corresponding damping ratio is
\begin{equation}
\zeta_I=\frac{b_0}{2\sqrt{JK}},
\label{eq-zeta-current}
\end{equation}
which gives $\zeta_I\simeq 0.15$. The ideal-current case is therefore strongly underdamped despite the intrinsic damping of the galvanometer. With zero initial conditions, its step response is
\begin{equation}
\theta_I(t)=\theta_{\mathrm{ss}}
\left[
1-e^{-\zeta_I\omega_n t}
\left(
\cos\omega_d t+
\frac{\zeta_I}{\sqrt{1-\zeta_I^2}}\sin\omega_d t
\right)
\right],
\label{eq-current-step-response}
\end{equation}
where
\begin{equation}
\omega_n=\sqrt{\frac{K}{J}},
\qquad
\omega_d=\omega_n\sqrt{1-\zeta_I^2}.
\label{eq-current-step-frequencies}
\end{equation}
We next drive the same galvanometer through the electrical port considered in Secs.~III and IV. The Thévenin source resistance is set equal to the external critical damping resistance, $R_s=R_{\mathrm{CDRX}}$, so that Eq.~(\ref{eq-critical-total-damping}) gives
\begin{equation}
b_0+\frac{k^2}{R_s+R_c}=2\sqrt{JK},
\end{equation}
and hence $\zeta=1$ in the quasi-static limit. Unlike the ideal-current case, the coil current is now determined dynamically by the applied voltage and by the motion-induced EMF. The numerical response is therefore obtained from the complete coupled equations, Eq.~(\ref{eq-coupled-time}), with $\tau=0$, together with the source relation, Eq.~(\ref{eq-source-termination}). This is a third-order system, with $\theta$, $\dot\theta$, and $i$ as its dynamical variables. No electromagnetic damping term is added explicitly: its effect arises from the electromechanical coupling and the electrical termination.

To make the comparison depend only on the transient dynamics, the voltage step is chosen to produce the same steady deflection $\theta_{\mathrm{ss}}=30^\circ$ as in the ideal-current case. Since the back EMF vanishes at steady state,
\begin{equation}
V_s=(R_s+R_c)I_0.
\label{eq-voltage-step-amplitude}
\end{equation}
This gives $V_s=8.46\,\mathrm{mV}$.

\begin{figure}[!ht]
\centering
\includegraphics[width=0.75\columnwidth]{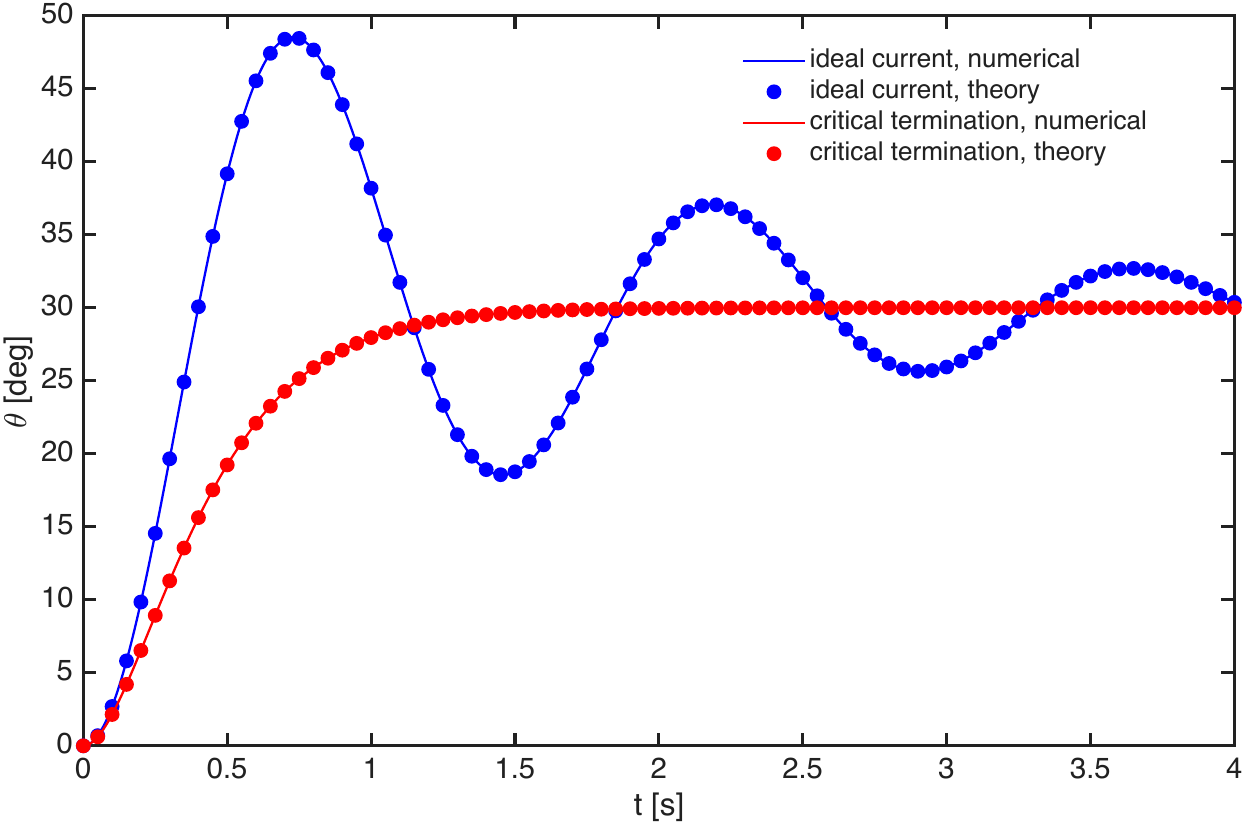}
\caption{Step responses of the example galvanometer for the two electrical driving conditions considered in the text, both producing the same steady deflection $\theta_{\mathrm{ss}}=30^\circ$.}
\label{fig-step-response}
\end{figure}

The complete third-order system is integrated from zero initial conditions and compared with the corresponding second-order prediction. For the critically terminated case, the quasi-static theory gives
\begin{equation}
\theta_C(t)=\theta_{\mathrm{ss}}
\left[1-(1+\omega_n t)e^{-\omega_n t}\right].
\label{eq-critical-step-response}
\end{equation}
The electrical relaxation time is $L/R=4.30\,\mu\mathrm{s}$, whereas the mechanical time scale is $1/\omega_n=0.229\,\mathrm{s}$, a separation by a factor of $5.33\times10^4$. Consistently with the criterion established in Sec.~IV, the complete third-order response differs from the quasi-static critical response by at most $4.01\times10^{-4}\%$ of the final deflection. Figure~\ref{fig-step-response} shows the two driving conditions together with their theoretical responses.

The expressions used to calculate $J$, $R_c$, and $L$, together with the MATLAB code (The MathWorks Inc., Natick, Massachusetts, 2026) that generates the numerical step responses of Fig.~\ref{fig-step-response}, are provided in Appendix~\ref{app-matlab}.

\section{Conclusion}

The moving-coil system considered in this paper provides a simple example of how energy exchange across an electromechanical interface can make the dynamics of a physical system depend on its electrical termination. Starting from the magnetic pendulum, the analysis shows how the same electromechanical coupling governs both the torque produced by the current and the back EMF generated by motion. In the radial-field geometry, this coupling becomes independent of angular position, making the bilateral character of the interaction particularly transparent.

This bilateral interaction is also the physical origin of electromagnetic damping. Motion generates an electrical response that modifies the current and hence the torque acting back on the mechanical system. The resulting feedback depends on the electrical termination: in the two-port representation, its effect appears as an electrical impedance reflected into the mechanical domain. The numerical example makes the consequence explicit: the same moving-coil system, driven to the same equilibrium position, can exhibit markedly different transient dynamics solely because its electrical boundary conditions have changed.

For measurement systems, in particular, this provides a useful reminder that a transducer and the circuit connected to it cannot always be treated as dynamically independent. Loading may affect not only the magnitude of a measured signal but also the dynamics through which that signal is produced. More generally, the example shows how apparently separate electrical and mechanical descriptions can be understood as different views of a single energy-exchanging system. The moving-coil system therefore offers an undergraduate setting in which familiar electromagnetism and oscillator dynamics lead naturally to a broader physical picture of how energy exchange shapes the dynamics of coupled systems.

\appendix
\section{Numerical implementation}
\label{app-matlab}

The following MATLAB code evaluates the geometrical parameters reported in Table~\ref{tab-prototype} and generates Fig.~\ref{fig-step-response}. The ideal-current case is integrated as the reduced second-order mechanical system, whereas the critically terminated case is integrated as the complete third-order electromechanical system. Forward Euler integration is used throughout. 

\begin{verbatim}
% d'Arsonval galvanometer: step responses
clear; close all; clc
% Geometry and material properties
N = 100;  a = 1e-2;  ell = 1.5708e-2;  r = 50e-6;
B = 0.2;  rho_m = 8960;  rho_e = 1.68e-8;  mu0 = 4*pi*1e-7;
w  = 2*a;
k  = 2*N*a*ell*B;
J  = 2*N*rho_m*pi*r^2*a^2*(ell + 2*a/3);
Rc = rho_e*N*(2*ell + 2*w)/(pi*r^2);
Lself = @(x) (mu0/(2*pi))*x*(log(2*x/r) - 1);
Mmut  = @(x,d) (mu0/(2*pi))* ...
    (x*asinh(x/d) - sqrt(x^2+d^2) + d);
L = N^2*(2*Lself(ell) + 2*Lself(w) ...
    - 2*Mmut(ell,w) - 2*Mmut(w,ell));
% Mechanical parameters and critical termination
K = 6.000e-7;
b0 = 4.170e-8;
Rs = k^2/(2*sqrt(J*K) - b0) - Rc;
R  = Rs + Rc;
% Inputs giving the same steady deflection
theta_ss = deg2rad(30);
I0 = K*theta_ss/k;
Vs = R*I0;
% Theoretical parameters
wn = sqrt(K/J);
zetaI = b0/(2*sqrt(J*K));
wd = wn*sqrt(1-zetaI^2);
% Forward Euler integration
h = 1e-6;
t = (0:h:4)';
n = numel(t);
% Ideal current drive: x = [theta; theta_dot]
xI = zeros(2,n);
for q = 1:n-1
    dx = [xI(2,q);
        (-K*xI(1,q) - b0*xI(2,q) + k*I0)/J];
    xI(:,q+1) = xI(:,q) + h*dx;
end
% Critical voltage drive: x = [theta; theta_dot; i]
xC = zeros(3,n);
for q = 1:n-1
    dx = [xC(2,q);
        (-K*xC(1,q) - b0*xC(2,q) + k*xC(3,q))/J;
        (Vs - R*xC(3,q) - k*xC(2,q))/L];
    xC(:,q+1) = xC(:,q) + h*dx;
end
% Theoretical second-order responses
thetaI = theta_ss*(1 - exp(-zetaI*wn*t).* ...
    (cos(wd*t) + zetaI/sqrt(1-zetaI^2)*sin(wd*t)));
thetaC = theta_ss*(1 - (1 + wn*t).*exp(-wn*t));
% Plot
q  = 1:2000:n;
q1 = 1:50000:n;
figure
plot(t(q),rad2deg(xI(1,q)),'b-', ...
    t(q1),rad2deg(thetaI(q1)),'b.', ...
    t(q),rad2deg(xC(1,q)),'r-', ...
    t(q1),rad2deg(thetaC(q1)),'r.','LineWidth', 1.1, 'MarkerSize', 18)
xlabel('t [s]')
ylabel('\theta [deg]')
legend('ideal current, numerical','ideal current, theory', ...
    'critical termination, numerical','critical termination, theory', ...
    'Location','best','Box','off')
set(gca, "FontSize", 14, "LineWidth", 1.5, "Box", "on")
% Numerical checks
bem = k^2/R;
zetaC = (b0 + bem)/(2*sqrt(J*K));
tau_e = L/R;
tau_m = 1/wn;
errI = max(abs(xI(1,:)' - thetaI))/theta_ss*100;
errC = max(abs(xC(1,:)' - thetaC))/theta_ss*100;
fprintf('I0       = %.6e A\n',I0)
fprintf('Vs       = %.6e V\n',Vs)
fprintf('zeta_I   = %.6f\n',zetaI)
fprintf('zeta_C   = %.6f\n',zetaC)
fprintf('b_em     = %.6e N m s\n',bem)
fprintf('L/R      = %.6e s\n',tau_e)
fprintf('1/wn     = %.6e s\n',tau_m)
fprintf('ratio    = %.3e\n',tau_m/tau_e)
fprintf('error I  = %.6f %%\n',errI)
fprintf('error C  = %.6f %%\n',errC)
\end{verbatim}

%
%






\end{document}